\documentclass[11pt,twoside]{article}
\usepackage{asp2004}
\usepackage{epsf}
\usepackage{psfig}
\usepackage{lscape}
\usepackage{graphicx} %package rajoute
\def\kms{km~s$^{-1}$}
\def\vsini{$V\!\sin i$}
\def\teff{$T_{\rm eff}$}
\def\logg{$\log~g$}
\def\omc{$\Omega/\Omega_{\rm{c}}$}
\def\ttms{$\frac{\tau}{\tau_{MS}}$}

\begin{document}
\title{Effects of metallicity, star formation conditions and evolution of B \&
Be stars.}   %%% Fill in title


\author{C. Martayan, A.-M. Hubert, M. Floquet, C. Neiner}   %%% Fill in author names
\affil{GEPI, UMR\,8111 du CNRS, Observatoire de Paris-Meudon, 92195 Meudon Cedex, France}    %%% Fill in author affiliations
\author{Y. Fr\'emat}   %%% Fill in author names
\affil{Royal Observatory of Belgium, 3 avenue circulaire, 1180 Brussels, Belgium}    %%% Fill in author affiliations
\author{J. Zorec}   %%% Fill in author names
\affil{Institut d'Astrophysique de Paris (IAP), 98bis boulevard Arago, 75014 Paris,
France}    %%% Fill in author affiliations

\begin{abstract} %%% Abstract to run on from here.

To study the effects of metallicity and evolution on the appearance 
of the Be phenomenon in the B stars population, we observed several 
fields in the Large and Small Magellanic Clouds (LMC and SMC, 
respectively) which have different metallicities. Thanks to the 
FLAMES-GIRAFFE multi-fibres spectrograph on the VLT-UT2, we obtained 
spectra of 520 stars in the LMC-NGC2004 and SMC-NGC330 regions. We 
used 2 settings at medium resolution: R=8600 for the red setting 
which contains H$\alpha$ and R=6400 for the blue setting which 
contains H$\gamma$, H$\delta$, \ion{He}{I} 4026, 4388, 4471 \AA. The 
latter setting was used to obtain fundamental parameters of the stars 
by fitting the observed spectrum with theoretical spectra. We used 
TLUSTY (Hubeny \& Lanz 1995) to compute a grid of model atmospheres 
with abundance adopted from Korn et al. (2002) for the LMC and from 
Jasniewicz \& Th\'evenin (1994) for the SMC. Thanks to the GIRFIT 
code (Fr\'emat et al. 2005a), we obtained the fundamental parameters 
\teff, \logg, \vsini~and radial velocity (RV) for each star of the 
samples.We took into account the effects of fast rotation (stellar 
flattening and gravitational darkening) for Be stars to correct their 
apparent fundamental parameters.

Then we compared the rotational velocities between fields and 
clusters in the SMC and in the LMC respectively, between the LMC and 
the SMC, and between the MC and the Milky Way (MW). The results show 
an increase in Vsini with decreasing metallicity in B and Be stars 
populations. The evolutionary status and ages of Be stars were also 
investigated.

\end{abstract}

%%% MAIN BODY OF TEXT GOES HERE. CONSULT "INSTRUCTIONS FOR AUTHORS USING
%%% LATEX2E MARKUP", SECTIONS 2.3-2.6 FOR HELP WITH EQUATIONS, FIGURES,
%%% AND TABLES.

%\section{}   %%% Top level section head (remove "%" symbol)
%\subsection{}   %%% Second level section head (remove "%" symbol)
%\subsubsection{}   %%% Lowest level section head (remove "%" symbol)
%\section*{}	%%% Unnumbered top level section head (remove "%" symbol)
%\subsection*{}   %%% Unnumbered second level section head (remove "%" symbol)

\section{Introduction and observations}
In the Milky Way (MW), the behaviour of B and Be stars has been investigated
for several decades and recently new results have been obtained. According to
Fabregat \& Torrej\'on (2000), the Be phase appears mainly in the second part
of the MS; however, we do not know whether only a fraction or all B-type 
stars will become Be stars. According to Maeder et al. (1999), the proportion
of Be stars increases with the decreasing metallicity of the environment (MW,
LMC and SMC).  Porter \& Rivinius (2003) have found that 1/3 of
Be stars are binaries at the maximum, then the binarity cannot explain the 
Be phenomenon in all cases.

In the Magellanic Clouds (MC) the B and Be star populations are not well known or
unknown and there is no spectral classification for these objects.\\

In order to investigate the effects of metallicity, star formation conditions
and evolution in B \& Be stars, we observed B-type stars, selected on
colour-magnitude criteria, with the VLT multifibres GIRAFFE spectrograph. These
observations were performed at medium resolution in 2 settings: a ``blue''
setting  centered at 4250 \AA~ to determine the fundamental parameters and  a
``red'' setting centered at 6570 \AA~ for the characterization of the emission of
Be stars. Finally, we observed 177 stars in the Large Magellanic Cloud
(LMC); among them, 25 are new Be stars, 22 are known Be stars and 121 stars are
B stars. In the Small Magellanic Cloud (SMC), we observed 346 stars, of which
90 are new Be stars, 41 are known Be stars and 202 are B stars.
In the statistical studies reported in the
following sections, we have removed from the samples
the binaries (B and Be stars) and we have assumed a random distribution of the 
inclination angle.

\section{Fundamental parameters determination}

We have determined the fundamental parameters (\teff, \logg, \vsini) with the
GIRFIT least squares procedure, which is able to handle large datasets and was
previously developed and described by Fr\'emat et al. (2005a). GIRFIT fits the
observations with theoretical spectra interpolated in a grid of stellar  fluxes
computed with the SYNSPEC programme and from model atmospheres calculated with
TLUSTY (Hubeny \& Lanz 1995, see references therein) or/and  with ATLAS9
(Kurucz 1993; Castelli et al. 1997) and with suitable abundances for the LMC
(Korn et al. 2002) and for the SMC (Jasniewicz \& Th\'evenin 1994). It accounts
for the instrumental resolution through convolution of spectra with a Gaussian
function and for Doppler broadening due to rotation. Use is made of subroutines
taken from the ROTINS computer code provided with SYNSPEC (Hubeny \& Lanz
1995). Thanks to calibration (Gray \& Corbally 1994 and Zorec 1986), we have
obtained the spectral classification of the stars. The main types range from B1
to B3. The luminosity classes are mainly V for LMC B stars and V \& IV for SMC
B stars.The luminosity classes are mainly III for LMC Be stars and IV \& III for
SMC Be stars. Consequently, the B stars in the SMC seem to be slightly evolved 
in comparison with the B stars in the LMC. The Be stars in the SMC and in the LMC
seem to be evolved. However, following Fabregat \& Gutierr\'ez-Soto (2004), there is 
no clear relation between the evolutionary status of a Be star and its luminosity class. 

By interpolation in theoretical HR tracks with Z=0.004 for the LMC 
(Charbonnel et al. 1993) and with Z=0.001 for the SMC (Schaller et al. 1992), we
have obtained other associated parameters like the luminosity, the radius, the mass
and the age of the stars.

\section{Effects of metallicity and ages on the linear rotational velocities}
In order to investigate the potential effect of metallicity on the rotational velocities, 
we have made sub-samples of stars by mass selection. It allows to compare the results directly 
with the existing theoretical evolutionary tracks.
We assume that the distribution of the inclination angle is random and we have removed all the detected binaries 
from the sub-samples. The comparisons given in the next sections are 
supported by statistical tests like the Student t-test.

%We have found that Be stars begin their Main Sequence (MS) with a greater rotational velocity than B stars. 

\subsection{B stars}

The results on the mean rotational velocities of B stars in the SMC and in the LMC are given in Table~\ref{vsiniBmasses}.
Only the sub-samples of the 5-10 M$_{\odot}$ category are comparable because they have similar masses and ages 
while their metallicity is different.
The values of the rotational velocities, 156 \kms~and 119 \kms~ for B stars in the SMC and in the LMC respectively,
show an unambiguous effect of the metallicity. The lower the metallicity is, the higher the rotational velocities are.

\begin{table*}[th]
\scriptsize{
\caption[]{Comparison by mass sub-samples of the mean rotational velocities in the SMC and LMC B stars.
For each sub-sample, the mean age, mean mass, mean \vsini~and the number of stars (N*) are given.}
\centering
\begin{tabular}{l|cccc|cccc|cccc}
\hline
\multicolumn{1}{c}{}&\multicolumn{4}{c}{2-5 M$_{\odot}$}&\multicolumn{4}{c}{5-10 M$_{\odot}$}\\
\hline
	& $<$age$>$ & $<$M/M$_{\odot}$$>$& $<$\vsini$>$ & N* & $<$age$>$ & $<$M/M$_{\odot}$$>$& $<$\vsini$>$ & N* \\
\hline
SMC B stars & 8.1& 4.0& 166 $\pm$ 20 & 116 & 7.7& 6.6& 156 $\pm$ 20 & 77 \\
LMC B stars & & & & & 7.6&  7.2& 119 $\pm$ 20 & 87 \\
\hline
\end{tabular}

\begin{tabular}{l|cccc}
\multicolumn{1}{c}{}&\multicolumn{4}{c}{10-15 M$_{\odot}$} \\
\hline
	& $<$age$>$ & $<$M/M$_{\odot}$$>$& $<$\vsini$>$ & N* \\
\hline
SMC B stars & & & & \\
LMC B stars & 7.2& 11.6& 116 $\pm$ 20 & 13 \\
\hline
\end{tabular}
\label{vsiniBmasses}
}
\end{table*}

\subsection{Be stars}

Like for B stars, we have compared Be stars in the MC and in the MW. 
The rotational velocities of MW Be stars come from Chauville et al. (2001) and their masses and ages from Zorec et al. (2005).
The results are shown in Table~\ref{vsiniBemasses}.
We can compare the Be stars with similar masses and ages, for example:
SMC and MW Be stars of the 2-5 M$_{\odot}$ category; SMC, LMC and MW Be stars of the 5-10 M$_{\odot}$ and 
10-12 M$_{\odot}$ categories. They show a similar effect of metallicity: the lower the metallicity is, 
the higher the rotational velocities are. We cannot compare directly the MC and the MW Be stars of the 12-18 M$_{\odot}$ 
category because there are no MW Be stars at the age of MC Be stars. The Be phase seems to be longer in the MC than in the MW.

\begin{table*}[!th]
\scriptsize{
\caption[]{Comparison by mass sub-samples of the mean rotational velocities in the SMC, LMC and in the MW Be stars.
The latter ones come from Chauville et al. (2001) and Zorec et al. (2005).
For each sub-sample, the mean age, mean mass, mean \vsini~and the number of stars (N*) are given.}
\centering
\begin{tabular}{l|cccc|cccc}
\hline
\multicolumn{1}{c}{}&\multicolumn{4}{c}{2-5 M$_{\odot}$}&\multicolumn{4}{c}{5-10 M$_{\odot}$}\\
\hline
	& $<$age$>$ &  $<$M/M$_{\odot}$$>$ & $<$\vsini$>$ & N* & $<$age$>$ & $<$M/M$_{\odot}$$>$ & $<$\vsini$>$ & N* \\
\hline
SMC Be stars & 8.3 &  3.8 & 277 $\pm$ 40 & 14 & 7.6 & 7.7 & 295 $\pm$ 40 & 82 \\
LMC Be stars &  & & & & 7.5 & 7.7 & 285 $\pm$ 30 & 21 \\
MW Be stars & 8.1 & 4.4 & 241 $\pm$ 10 & 18 & 7.4 & 7.3 & 234 $\pm$ 10 & 52 \\ 
\hline
\end{tabular}

\begin{tabular}{l|cccc|cccc}
\multicolumn{1}{c}{}&\multicolumn{4}{c}{10-12 M$_{\odot}$}&\multicolumn{4}{c}{12-18 M$_{\odot}$}\\
\hline
	& $<$age$>$ & $<$M/M$_{\odot}$$>$ & $<$\vsini$>$ & N*& $<$age$>$ & $<$M/M$_{\odot}$$>$ & $<$\vsini$>$ & N* \\
\hline
SMC Be stars & 7.3 & 10.9 & 345 $\pm$ 40 & 13 & 7.2 & 13.5 & 324 $\pm$ 40 & 13\\
LMC Be stars & 7.3 & 11 & 259 $\pm$ 30 & 13 & 7.1 & 14.6 & 224 $\pm$ 30 & 10\\
MW Be stars & 7.1 & 10.6 & 231 $\pm$ 10 & 9 & 6.6 & 14.9 & 278 $\pm$ 10 & 16\\
\hline
\end{tabular}

\label{vsiniBemasses}
}
\end{table*}

\subsection{ZAMS rotational velocities}

Thanks to theoretical evolutionary tracks interpolated in Meynet \& Maeder (2000, 2002), Maeder \& Meynet (2001)  
and thanks to those calculated by Zorec et al. (2005 in preparation), we have interpolated the rotational 
velocities at the ZAMS for the Be stars in the MC (Martayan et al. 2005c in preparation) and in the MW.

\begin{figure}[!htbp]
\centering
\includegraphics[width=5 cm, height= 11cm,angle=-90]{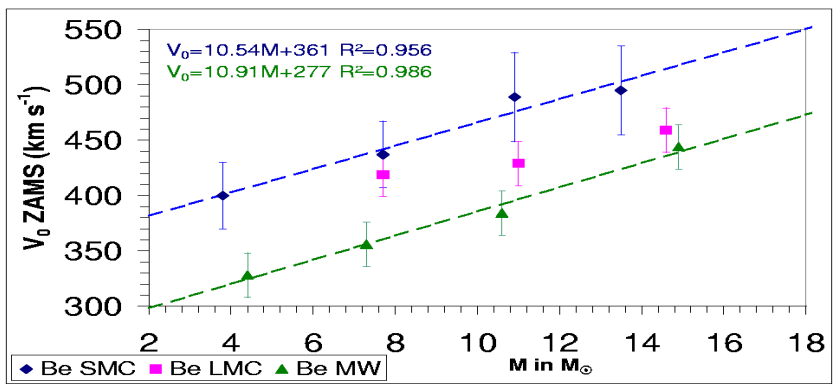}%V0ZAMSequawind.ps}%width=9 cm, height= 17cm,
\caption[]{ZAMS rotational velocities of Be stars in the SMC (blue diamonds), in the LMC (pink squares)
and in the MW (green triangles). the lines correspond to the linear regressions. 
Their corresponding equations and correlation coefficient are given in the upper left corner.
}
\label{V0ZAMSequawind}
\end{figure}

The results are shown in Figure~\ref{V0ZAMSequawind} and consequently, we note that: 
\begin{itemize}
\item The ZAMS rotational velocities depend on the mass of the stars and they present the same tendency in the SMC and in the MW.
\item The lower the metallicity is, the higher the ZAMS rotational velocities are.
\item At a given metallicity, there is a minimum value of mean ZAMS rotational velocities for 
which B stars will become or not Be stars. 
\end{itemize}

\section{Angular velocity and evolution}

With formula given in Chauville et al. (2001) and in Martayan et al. (2005a), 
we have obtained the ratio of the mean angular velocity to the breakup
velocity (\omc) for each sub-sample of Be stars in the MC.
The results are given in Table~\ref{omegaMC} and they show:
\begin{itemize}
\item In average, all Be stars have a ratio greater than \omc=70\%.
\item The tendency is similar in the LMC and in the MW (Cranmer 2005)
while it is different in the SMC.
\item Less massive stars in the SMC are close to the breakup velocity.
\item Massive Be stars in the SMC, which are evolved, seem to be critical rotators.
Contrary to the MW, the more massive Be stars can become critical rotators in a 
very low metallicity environment like in the SMC. This result is in agreement with the 
theoretical results of Meynet \& Maeder (2000) and Maeder \& Meynet (2001).
The results of the comparisons will be published in Martayan et al. (2005b in preparation).
\end{itemize}

\begin{table*}[th]
\centering
%\footnotesize
\small{
\caption[]{Results of the ratios of linear and angular mean velocities 
to the breakup velocities for the Be stars in the SMC and in the LMC.}
\centering
\begin{tabular}{@{\ }c@{\ \ \ }c@{\ \ \ }c@{\ \ \ }c@{\ \ \ }c@{\ \ \ }c@{\ \ \ }c@{\ \ \ }c@{\ \ \ }c@{\ \ \ }c@{\ }}
\hline
\hline
 & & LMC Be & & Z=0.004\\
\hline
$<$M/M$_{\odot}$$>$ & $<$R/R$_{\odot}$$>$ & $<$\vsini$>$ & $\frac{V_{e}}{V_{c}}$ & \omc \\
\hline
7.7 & 5.8 & 285 & 0.72 & 0.85 $\pm$ 0.13 \\
11.0 & 9.3 & 259 & 0.69 & 0.83 $\pm$ 0.08 \\
14.6 & 12.7 & 224 & 0.61 & 0.75 $\pm$ 0.10 \\
\hline
 & & SMC Be  & & Z=0.001 \\
\hline
$<$M/M$_{\odot}$$>$ & $<$R/R$_{\odot}$$>$ & $<$\vsini$>$ & $\frac{V_{e}}{V_{c}}$ & \omc \\
\hline
3.8 & 4.1 & 277 & 0.85 & 0.94 $\pm$ 0.11 \\
7.6 & 7.9 & 295 & 0.87 & 0.95 $\pm$ 0.11 \\
10.9 & 15.5 & 345 & 0.99 & crit $\pm$ 0.12 \\
13.3 & 18.0 & 324 & 0.96 & 0.99 $\pm$ 0.14\\ 
\hline
\end{tabular}
\label{omegaMC}
}
\end{table*}

\section{Evolutionary status of Be stars in the MW, the LMC and the SMC}

Following Fabregat \& Torrej\'on (2000), the Be phase appears in the second part of the MS (\ttms$>$0.5) in the MW (i.e. log(t)$>$7).
According to Zorec et al. (2005) and Fr\'emat et al. (2005b), the effects of fast rotation appear for \omc$>$60\% while Be stars have
ratios of \omc$>$80\% in average in all galaxies. Then we must take into account the effects of fast rotation to determine the ages of
the stars. We have used the FASTROT code described in Fr\'emat et al. (2005b) and we have used ratios: \omc=85\% for Be stars in the LMC and
\omc=95\% for Be stars in the SMC. We have obtained a redistribution from the later to earlier spectral types and a redistribution from the
luminosity classes III, IV to IV, V. Then Be stars are less evolved what the ``apparent'' classification shows.

In the MW, Zorec et al. (2005) have obtained the evolutionary status of Be stars. They note that the massive Be stars are mainly not evolved
(\ttms$<$0.5) while the other Be stars are mainly evolved.

\begin{figure}[!h]
\centering
\includegraphics[width=5cm, height=11cm,angle=-90]{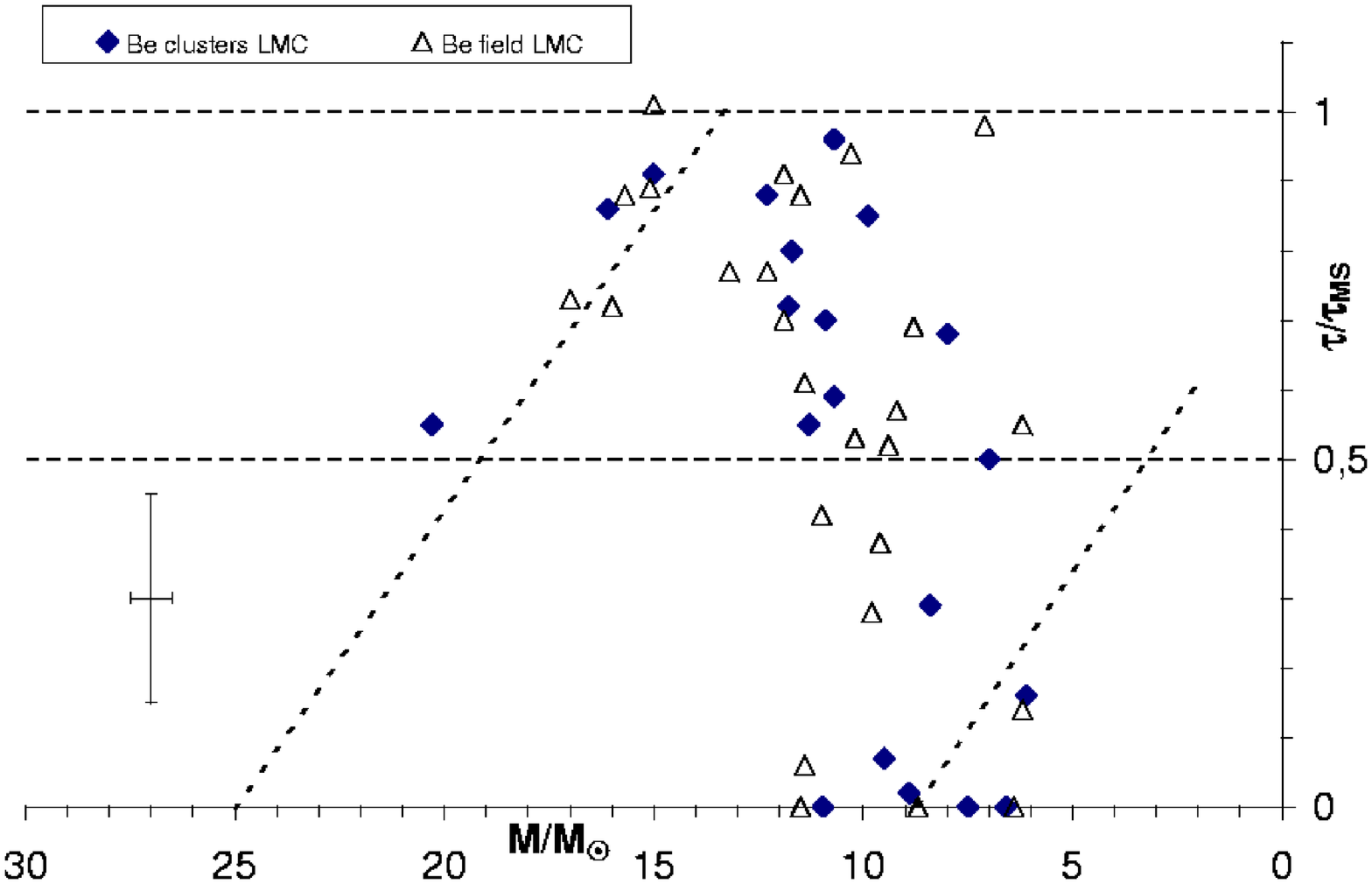}\\%evolBeLMCdroiteswind.ps} 
\includegraphics[width=5cm, height=11cm,angle=-90]{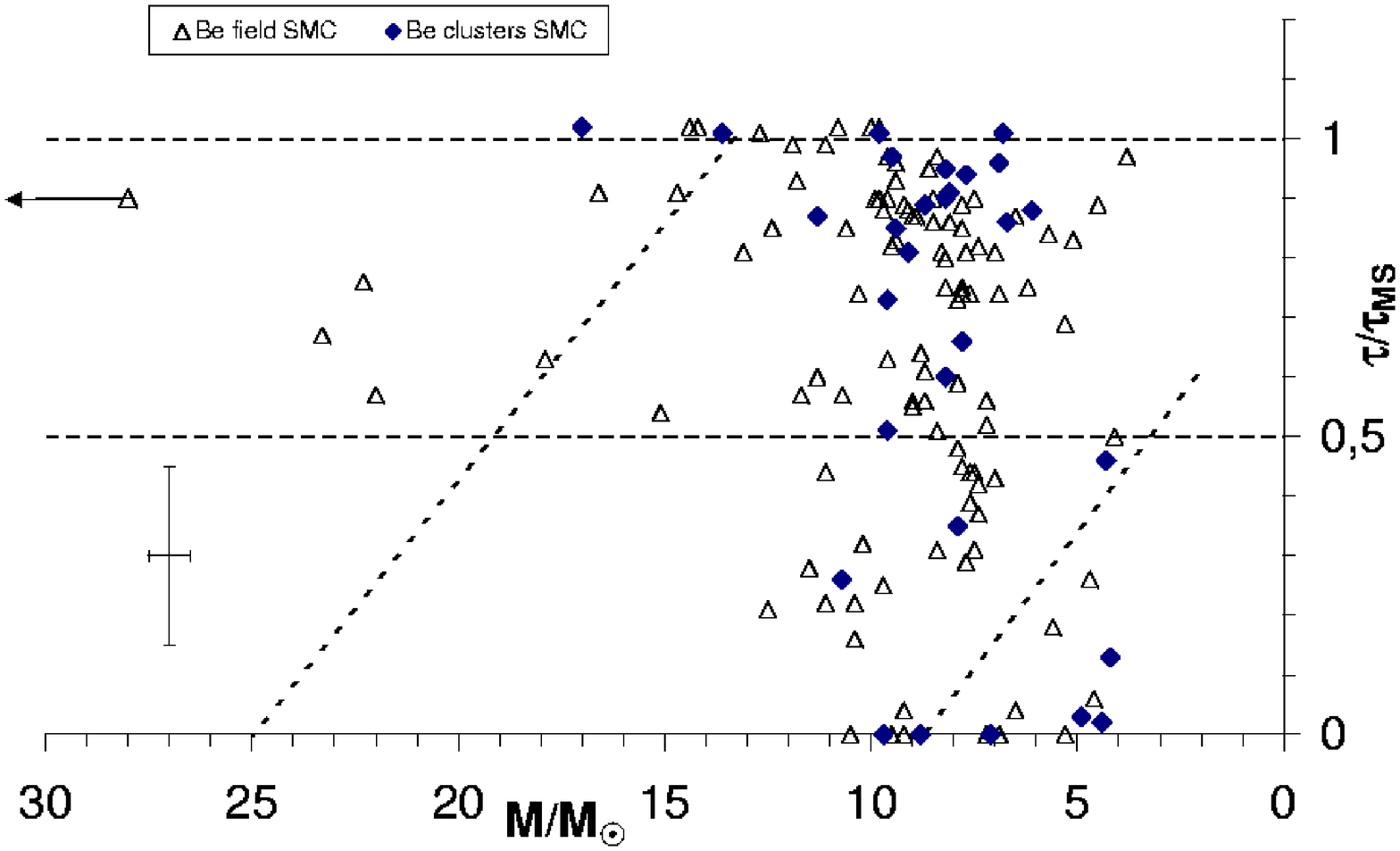}%evolBeSMCdroiteswind.ps}
\caption[]{Higher panel: evolutionary status of Be stars in the LMC. The fast rotation effects are taken into account with \omc=85\%.
Lower panel: evolutionary status of Be stars in the SMC. The fast rotation effects are taken into account with \omc=95\%.
In common: The typical errors are shown in the lower left corner. The diagonals come from Zorec et al. (2005) and show the area of existing
Be stars in the MW. The blue diamonds are for Be stars in the clusters and the triangles are for Be stars in the fields.
}
\label{statutevolBeMC}
\end{figure}

We have obtained in the same way, the evolutionary status of Be stars in the LMC (Martayan et al. 2005a) and in the SMC (Martayan et al.
2005b in preparation). 
The results are given in Figure~\ref{statutevolBeMC}.
The diagonals in this Figure come from Zorec et al. (2005) and show the area of existing Be stars in the MW. 
We note:
\begin{itemize} 
\item Contrary to the MW, massive Be stars will appear in the second part of the MS (\ttms$>$0.5).
\item Less massive Be stars are mainly evolved in agreement with Fabregat \& Torrej\'on (2000).
\end{itemize}

\section{Conclusions}

In conclusion, we have found an effect of metallicity on the rotational velocities for B and Be stars:
the lower the metallicity is, the higher the rotational velocities are. We have also found a similar effect of  
metallicity on the ZAMS rotational velocities for the progenitors of Be stars.
We have also found that Be stars begin their MS with a stronger rotational velocity than B stars. 
Then only a fraction of B stars with a sufficient initial rotational velocity will become Be stars.

Evolved massive Be stars seem to be critical rotators in the SMC.
Be stars seem to be mainly evolved even after the fast rotation corrections.
Contrary to the Milky Way, in the Magellanic Clouds the Be phase of massive stars appears in the second part 
of the Main Sequence.

%\acknowledgements %%% Text of acknowledgements runs on after this command.

%%% THE BIBLIOGRAPHY
%%%
%%% CONSULT SECTION 3 OF "INSTRUCTIONS FOR AUTHORS" FOR HOW TO USE NATBIB.
%%% AUTHORS ARE ENCOURAGED TO USE EITHER THE "THEBIBLIOGRAPY" ENVIRONMENT
%%% BY UNCOMMENTING (DELETING THE "%" SYMBOL) THE COMMANDS BELOW, OR BY
%%% USING THE BIBTEX ENVIRONMENT. TO FIND OUT WHICH IS APPLICABLE TO YOUR
%%% CONTRIBUTION, CONSULT THE VOLUME EDITORS FOR YOUR PROCEEDINGS.
%%%

\section*{Questions}
\textbf{Dr. D. Baade}: The well-known intermittency of the Be phenomenon, which seems
to be quite different for early and late spectral subtypes, may introduce a bias 
into number distributions of Be stars with, e. g., age or metallicity.
Have you tried to correct for such a bias?\\
\textbf{C. Martayan}: We have not tried to correct this possible bias because we have mainly 
early-type stars in our samples. Moreover thanks to a cross-correlation with the MACHO database, 
we have obtained the light-curves for the quasi-totality of the stars in our samples and none of B 
stars shows a behaviour of a Be star. \\
\\
\textbf{Dr. H. Henrichs}: Did you determine the metallicity in your Be stars sample?\\
\textbf{C. Martayan}: We have used the metallicity of Korn et al. (2002) for the stars 
in the LMC and the metallicity from Jasniewicz \& Th\'evenin (1994) for the stars in the SMC.
The determination of abundances is difficult with the VLT-GIRAFFE spectra, which are not adapted:
low resolution, low signal to noise. However, simultaneously to GIRAFFE spectra, we have 
also obtained spectra with the high-resolution multifibres VLT-UVES for stars in the 
LMC and in the SMC. In the next months, we will determine the abundances of these stars.\\

\end{document}